\documentclass[prb,12pt,aps]{revtex4}
\usepackage[dvips]{graphics}
\pdfoutput=1 
\usepackage{bm}
\usepackage[usenames]{color}
\usepackage{rotating}
\usepackage[colorlinks=true]{hyperref}
\usepackage{mathtools}
\usepackage{tabularx}
\usepackage{tabu}
\usepackage[english]{babel}
\usepackage{fontenc}
\usepackage{graphicx}
\usepackage{setspace}
\newcommand{\ai}{{\it ab initio}}
\newcommand{\ea}{{\it et al.}}

\begin {document}

\title{First Principles Predictions of Superconductivity in Doped Stanene}
\author{Yusuf Shaidu and Omololu Akin-Ojo}
\affiliation{Theoretical Physics Department, African University of
Science and Technology, Km 10 Airport Road, Galadimawa, Abuja, Nigeria}
\date{\today}

\newpage

\begin{abstract}
Stanene, composed of tin atoms arranged in a single layer, is the tin
analogue of graphene and past studies predicted it to be a topological
insulator. An energy band gap (of $\sim 0.1$~eV) was obtained in 
previous calculations for the buckled honeycomb structure of stanene and, thus, 
 phonon-mediated superconductivity in this material is ruled out.
 In this work we investigated, from first principles calculations
within density functional theory (DFT), the possibility of producing
 phonon-mediated superconductivity in stanene by doping the material. 
It was found that doping with calcium
(lithium) leads to superconductivity, albeit, with a very 
low superconducting
transition temperature $T_c$ of $\sim 0.7$~K ($\sim 1.3$~K), even lower than
the value ($3.7$~K) for bulk $\beta$-Tin.
\end{abstract}

\maketitle

\clearpage
\makeatletter

\section {INTRODUCTION} \label{sIntroA}
Stanene is a two dimensional sheet composed of tin (Sn) atoms arranged in 
a honeycomb lattice, similar to graphene. 
\emph{Ab initio} calculations~\cite{SME09} indicate that, unlike graphene, 
a low-buckled configuration of stanene is more stable compared to the
planar geometry.  
This is as a result of the weak $\pi-\pi$ bonding between the atoms. The buckling enhances the overlap between $\pi$ and $\sigma$ orbitals
and stabilizes the system.~\cite{SZ13}
 Theoretical studies indicate that free standing stanene is a zero band gap
 semiconductor in the absence of spin orbit coupling (SOC). The
inclusion of SOC leads to a band gap of about 0.1eV.~\cite{SZ13,BB14}
Chemical functionalization has been
predicted~\cite{SZ13,BB14} to be capable of increasing the band gap to 
 up to $0.35$~eV. With this capability of tuning its bandgap,
stanene could be useful as a semiconductor in low-power-consumption
electronics. 

 From the predicted electronic zero/non-zero band gap of stanene, one
can conclude that phonon-mediated superconductivity is not possible in
this material, just like its carbon counterpart
graphene is not superconducting.  Recently, some 
calculations~\cite{ER14} have, however, shown that superconductivity can be induced in
graphene by n-doping at a carrier density exceeding $10^{15}$~cm$^{-2}$.
 Superconductivity has been produced in graphene by 
doping its surface with alkaline metal ad atoms.~\cite{PG12,CZWF13} This is reminiscent of
the induction of superconductivity in graphite intercalated compounds
(GICs).~\cite{PG12} Specifically, it was found that lithium-covered 
graphene has a higher superconducting transition temperature $T_c$ 
than its calcium-covered counterpart unlike the situation in GICs where
the opposite is the case.~\cite{CZWF13} 
 Yang \ea\ showed that the 
superconductivity in GICs is as a result of the interaction between the 
interlayer (IL) states and the $\pi^*$ states, and that, if an ad atom
superlattice could be created with graphene, a similar
situation could lead to superconductivity.~\cite{YS14}  This was confirmed experimentally for doped
graphene in Ref.~\onlinecite{PG12}.

With the electronic similarities between graphene and stanene, namely,
the existence of a Dirac cone
at the so called K point and the presence of a tunable band gap~\cite{SZ13,BB14}, could it be
possible to produce superconductivity in stanene in a manner similar to
the induction  of superconductivity in graphene? Doped stanene might
superconduct at an even higher transition temperature $T_c$ than doped
graphene owing to the fact that bulk tin (white Tin) is superconducting
at low temperatures without
doping, in contrast to bulk graphene (graphite) which is not
superconducting. Graphite can mainly be made superconducting 
by creating interlayer states in it via metal intercalation.

To this end we examine, from first principles calculations, within Density Functional Theory (DFT), 
the possibility of inducing superconductivity
in stanene by creating IL states in it by doping. We doped the surface
of stanene with an alkaline and an alkaline earth metal, 
lithium and calcium, respectively and predicted $T_c$ for these doped
systems. We note that decorated stanene (SnX, with X= OH, I, Cl, Br)
have been predicted to be topological superconductors when doped with silver
atoms~\cite{DecoratedStanene} but the superconducting property of SnX
itself was not predicted but only assumed in that study. 

 Stanene has recently been synthesized.~\cite{ExptStanene} It was
fabricated on a Bi$_2$Te$_3$(111)
substrate by the process of molecular beam epitaxy.~\cite{ExptStanene}
This experimental realization of stanene should open up the possibility of verifying
the different predictions made for the material.   

\section{METHODS}\label{sMethods}
We first determined the optimum structure of stanene and computed its
electronic band structure.  
The calculations were performed using Density Functional Theory (DFT) with the gradient corrected
exchange correlation functional of Perdew, Burke, and Ernzerhof (PBE).~\cite{PBE} The  
QUANTUM ESPRESSO \cite{QE09} suite of codes was used. 
Plane waves with a kinetic energy cutoff of 55~Ry were used in
expanding the wavefunctions and core electrons were excluded through
the use of a norm-conserving pseudopotential.~\cite{NC}
We inserted a vacuum layer with a width larger than $20$\AA\ in 
the non-periodic direction of the system.
Since the material under consideration is a semimetal,
 the electronic properties were determined by integration in the first
Brillouin zone (BZ)  
 using a Gaussian smearing having a width of
0.05~Ry.
 All configurations and their respective atomic positions were optimized ensuring 
that the maximum force on any atom was smaller than $10^{-4}$eV/\AA. 
The electronic band structure was calculated along the path connecting
the special high symmetry k-points 
 $\Gamma, M, K$ in the first BZ as shown
in Sect.~\ref{sResults} below. The electronic density of states (EDOS)
was computed using a finer mesh of the first BZ with electron momentum of $27\times27\times1$ and $21\times21\times1$ for 
stanene, and doped stanene respectively.

Next, the phonon band structure and phonon density of states (PDOS) were calculated in the framework of
density functional perturbation theory (DFPT) as implemented in the
QUANTUM ESPRESSO codes.~\cite{SB} These
calculations showed, among other things, the stability of the buckled
stanene structure. The electron-phonon coupling (EPC) 
constants $\lambda$ were also determined but only for doped stanene
 since undoped stanene does not superconduct. 
The necessary summations over the electron momenta and phonon
momenta (see Eqs.~\ref{Eliashberg} and \ref{PDOS} below) in computing the EPC constants were carried
out using an electron momentum mesh of $32\times32\times1$ and a phonon 
momentum one of $8\times8\times1$. 

We used the following formula by Allen and Dynes~\cite{AD75} to determine the 
 superconducting transition temperature $T_c$:
\begin{equation}
 T_c=\frac{
\omega_{log}}{1.20}\exp\left(-\frac{1.04(1+\lambda)}{\lambda-\mu^*(1+0.62\lambda)}\right), \label{AD}
\end{equation}

where, $\omega_{log}$, $\lambda$ and $\mu^*$ are the logarithmic average of the phonon energy, electron-phonon coupling constant
and the electron-electron Coulomb repulsion parameter, respectively.
 The first two quantities were calculated as:
\begin{equation}
\omega_{log}=\exp\left(\frac{2}{\lambda}\int_{0}^{\infty}\alpha^2F(\omega)\ln{\omega}\frac{\mathrm{d\omega}}{\omega}\right) \label{wlog}
\end{equation}

\begin{equation}
\lambda=2\int_0^{\infty}\frac{\alpha^2(\omega)F(\omega)}{\omega}\mathrm{d\omega}
\label{lambda}
\end{equation}

where, the Eliashberg function $\alpha^2F$ is given by
\begin{equation}
 \alpha^2F(\omega)=\frac{1}{N(0)N_{\bf k}N_{\bf k'}}\sum_{{\bf k,\bf
k'},v}|M_{\bf k,\bf k'}^{\nu}|^2\delta(\epsilon_{\bf
k})\delta(\epsilon_{\bf k'})\delta(\omega-\omega_{ \bf q,\nu}). 
\label{Eliashberg}
\end{equation}

 The terms that appear in the expression are defined as follows: $N(0)$
is the electronic density of states per spin at the Fermi level, i.e., 
$N(0)=\sum_{\bf k} \delta(\epsilon_{\bf k})$ and  $N_{\bf k}$ and $N_{\bf q}$ represent the
total number of $\bf k$ and $\bf q$ points respectively. The Kohn-Sham
eigenvalue with reference to the Fermi level is $\epsilon_k$ and $M_{k,k'}$ is the electron-phonon coupling matrix element. The 
 Dirac delta distributions $\delta(\epsilon_{\bf k})$ and
$\delta(\epsilon_{\bf k'})$ restrict the electrons to the Fermi surface,
$\omega_{{\bf q},\nu}$ is the phonon frequency at the phonon wave vector
$\bf q$ of 
 the phonon branch of index $\nu$. Here, $k$ and $k'$ indicate both the electron wave vector and band index, and spin index. {$\bf{q}=\bf{k'-k}$}
indicates both the phonon wave vector and band index. 
In our calculations the Dirac delta distribution of Eq.~\ref{Eliashberg}
above was replaced by a Gaussian broadening function having a width of 0.01~Ry.  
 The variable $\alpha^2$ is the coupling strength 
and $F(\omega)$ is the phonon density of state, i.e., 
\begin{equation}
F(\omega)=\sum_{\bf q}\delta(\omega-\omega_{\bf q})\label{PDOS}
\end{equation}

\section{RESULTS AND DISCUSSION}\label{sResults}
The optimized lattice parameters are shown in Table~\ref{ST} along with
the results obtained from calculations in the Literature.
The variable $a$ is the lattice constant, $d$ is the out-of-plane
buckled height, and $a_{sn-sn}$ is the Sn--Sn bond length. The values
calculated in this work are consistent with other theoretical
predictions in the literature. We also found the buckled stanene
structure to be stable. The computed phonon band structure of pristine
stanene had no imaginary frequency. This stability is in agreement with Ref.~\onlinecite{BB14,SZ13} but
differs from the results presented in the supplementary information of
Ref.~\onlinecite{UnstableStanene}. 

\begin{figure}[h!]
\begin{center}
\rotatebox{0}{\scalebox{0.5}{\includegraphics{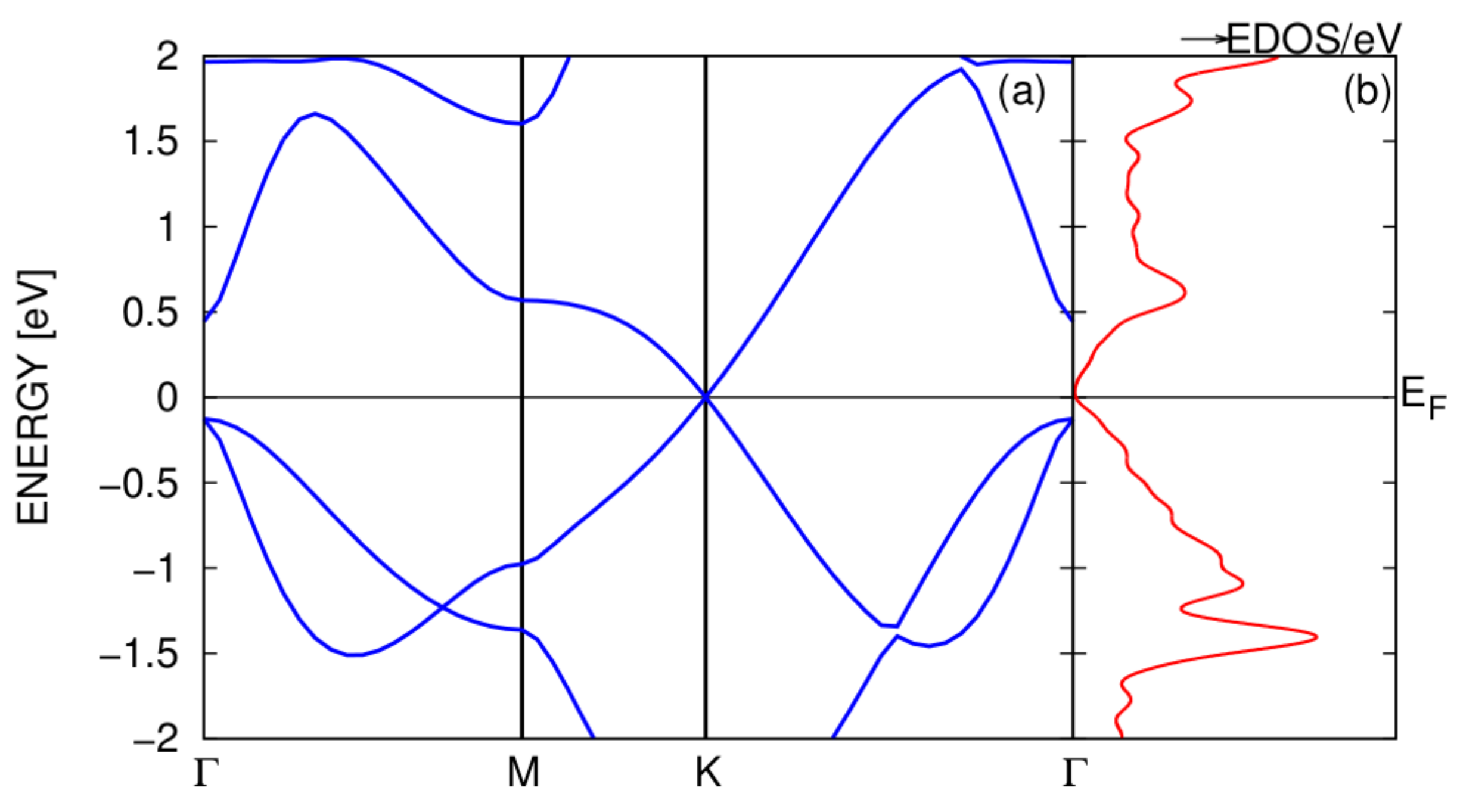}}}
\end{center}
\caption[Electronic band structure and EDOS of stanene]{{\bf (a)}, is
the electronic band structure and, {\bf (b)}, is the electronic density
of states (EDOS) of stanene, $E_F$ is the Fermi energy,
which we set as the reference level, and  $\Gamma (0,0,0), M(\frac{1}{2},0,0) ,\text{and } K (\frac{1}{3},\frac{1}{3},0)$ are 
special symmetry k points.}
\label{EBS}
\end{figure}

Next, we present the electronic band structure and the electronic
density of states of free standing stanene. 
The band structure and the density of states in Fig.\ref{EBS} are in good agreement with 
the results of Broek \ea~\cite{BB14} The figure
clearly shows that stanene has a Dirac point at K similar to the case in graphene. 
Thus, it is a zero band gap material. 
Our calculations did not include spin-orbit coupling (SOC) effects. It
is known that SOC opens up the gap by about 0.1~eV at the K point.
In either case, with or without SOC, the electron density at the Fermi level is very low for
stanene. Consequently, free-standing stanene
cannot superconduct due to a lack of carriers at the Fermi
level that can participate in Cooper pairing.

A first attempt towards making stanene a superconductor is to assume phonon mediated superconductivity and mimic the way in
which its two dimensional counterpart, graphene
was made superconducting as reported in Refs.~\onlinecite{YS14, PG12, F14}. 
Hence we doped stanene with an electron donor, 
in this case with Lithium and Calcium, respectively, in order to create
an electron rich Fermi surface. 
Electrons at the Fermi surface can then form Cooper pairs and make the
material superconducting. 

\begin{figure}[h!]
 \begin{center}
\rotatebox{0}{\scalebox{0.5}{\includegraphics{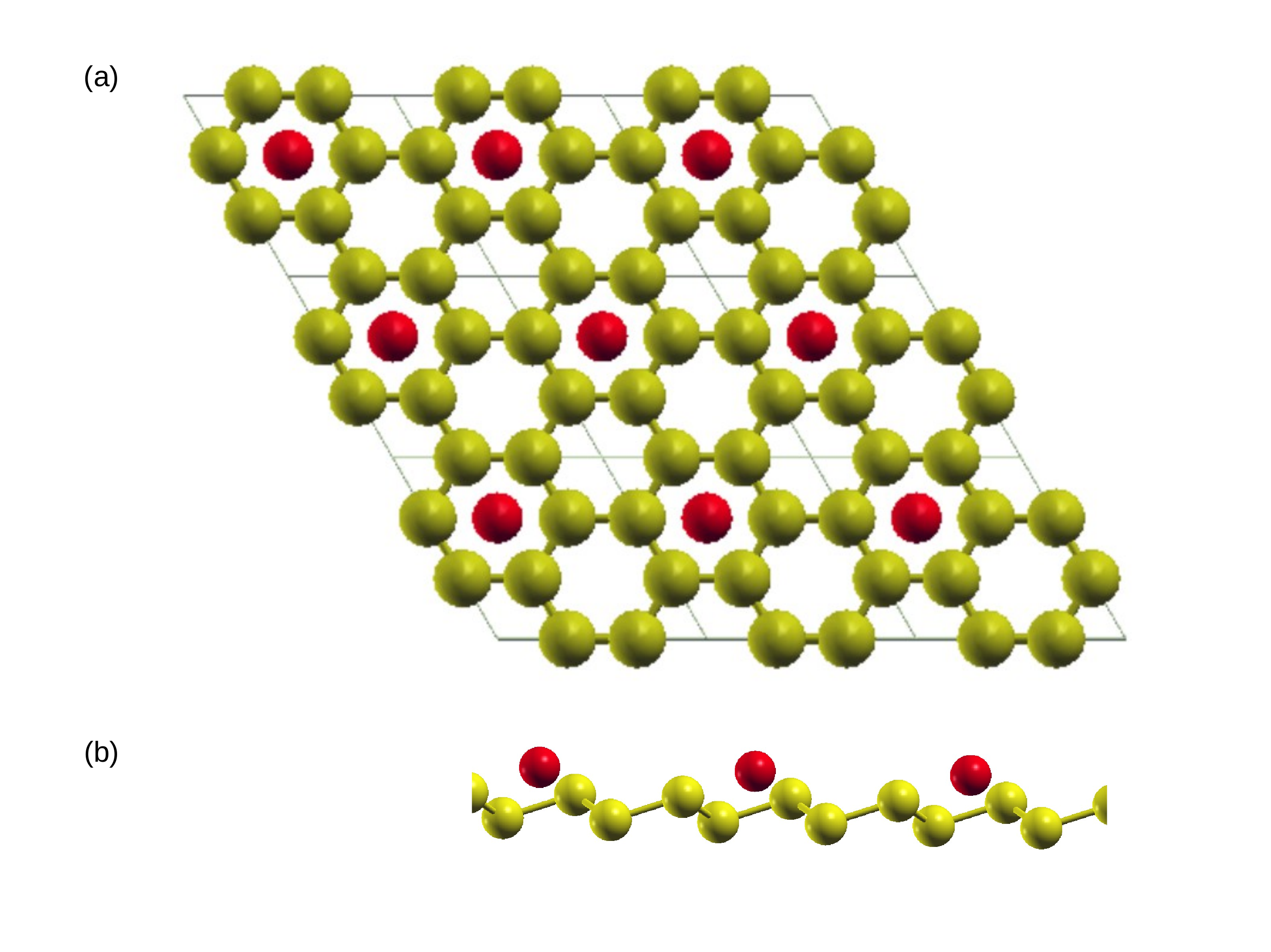}}}
\end{center}
\caption{Crystal structure of doped stanene. The yellow spheres
represent tin atoms while the red spheres are the dopants. Both the top
(a) and the side views (b) are shown above and below, respectively.}
\label{LSn}
\end{figure}

\begin{figure}[h!]
\centering
\rotatebox{0}{\scalebox{0.5}{\includegraphics{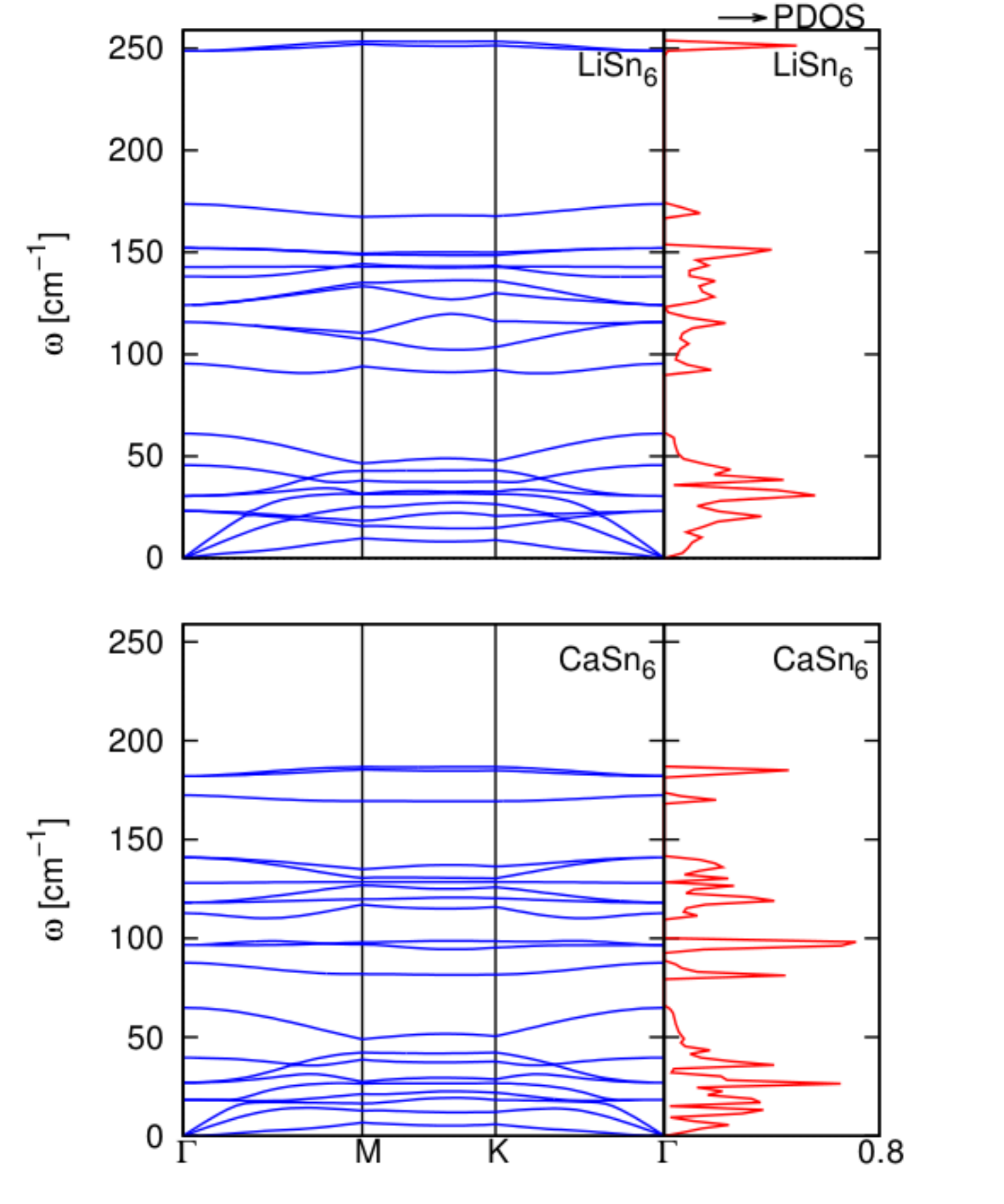}}}
\caption{Phonon dispersion (left) and the 
phonon density of states (right) for Lithium-doped stanene, LiSn$_6$
(top) and for Calcium-doped stanene, CaSn$_6$ (bottom).
The phonon dispersion are plotted along  $\Gamma,~M  ~K \text{and}~
\Gamma$ high symmetry k points of a 2D honeycomb crystal}\label{PLCS}
\end{figure}

We placed Lithium atoms and Calcium atoms in turn on stanene as shown in
Fig.~\ref{LSn} and optimized the resulting structures but with the fixed lattice constant
($a=4.64$~\AA) obtained for free-standing stanene. 
The optimized structure of Lithium (Calcium)  doped stanene reveals that the structure is stable when 
Lithium (Calcium) atoms were placed at $1.75~(2.24)$~\AA\  above the
stanene plane. No imaginary frequencies were seen in the computed phonon
dispersion curves (see Fig.~\ref{PLCS}).

\begin{figure}[h!]
\centering
\rotatebox{0}{\scalebox{0.5}{\includegraphics{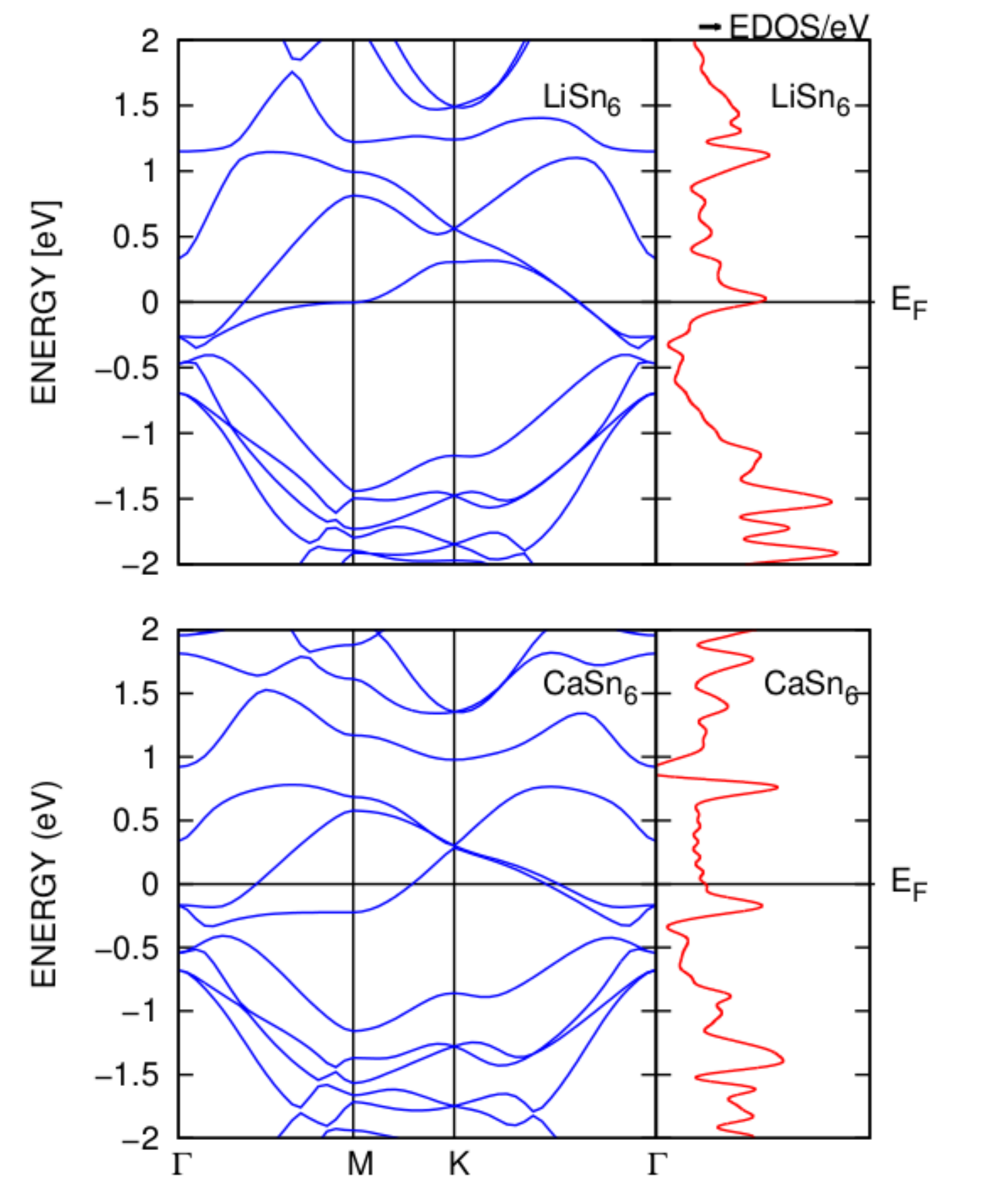}}}
\caption{Electronic Band Structures and the Total Density of States. The
top panel is for LiSn$_6$ and the bottom one for CaSn$_6$.
In each case, the electronic band structure is on the left while the
total electronic density of states (EDOS) is on the right hand side. $E_F$ is the Fermi energy,
which we set as the reference level with a horizontal line. The
band structure is plotted along the k-path connecting the $\Gamma,~M  ~K \text{and}~ \Gamma$ 
high symmetry k points of a 2D honeycomb crystal}\label{ELCS}
\end{figure}

The electronic band structures of Lithium-doped and Calcium-doped
stanene are shown in Fig.~\ref{ELCS}.  
 From the figure, it is clear that carriers are now available near the Fermi
surface in contrast to the situation with free-standing stanene
(Fig.~\ref{EBS}). 

\subsection{Eliashberg spectral function and electron-phonon coupling
constant}\label{ssEliashberg}
\begin{figure}[h!]
\centering
\rotatebox{0}{\scalebox{0.5}{\includegraphics{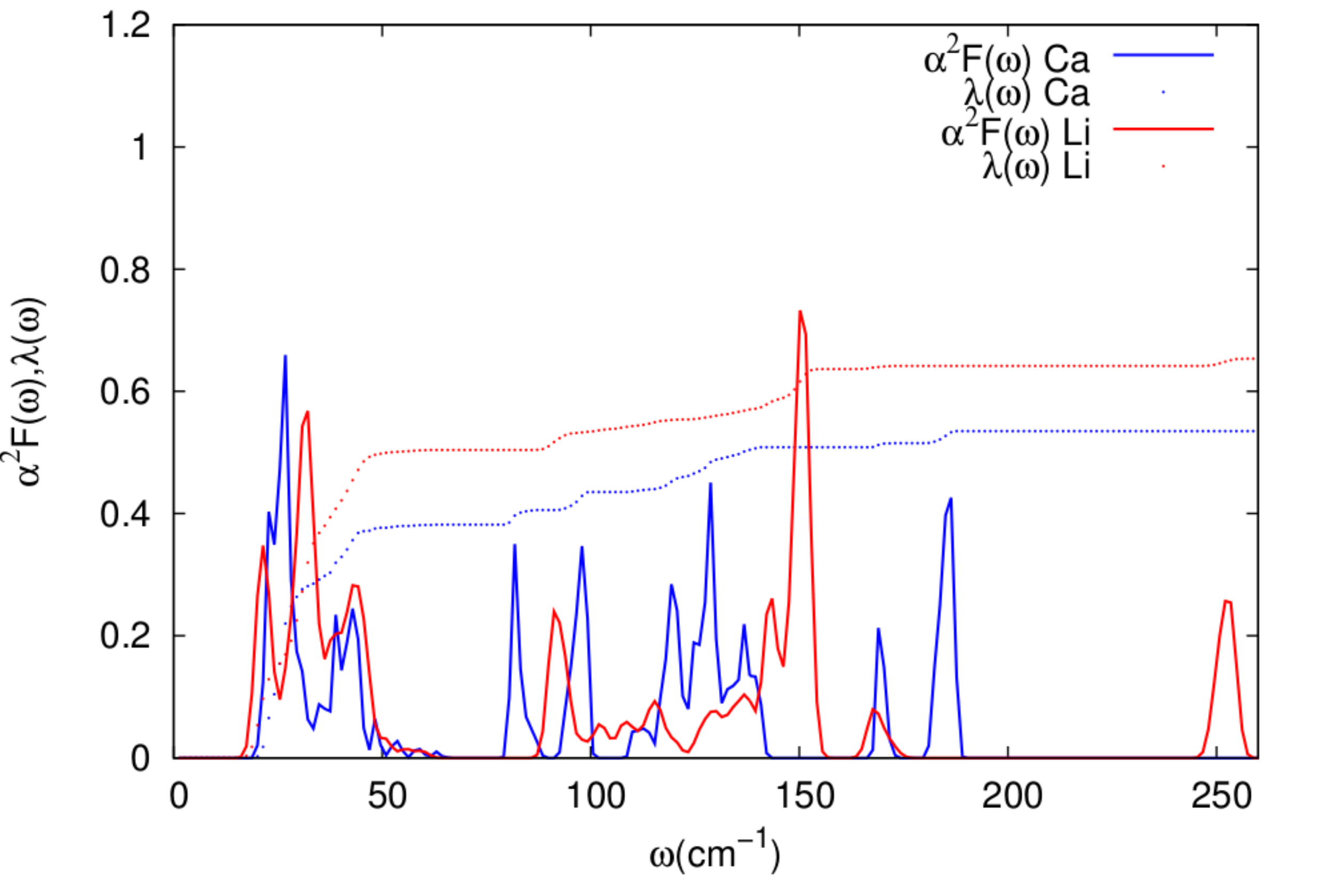}}}
 \vskip +1cm
 \caption{The Eliashberg spectral function $\alpha^2F(\omega)$ (solid
lines) and the
electron-phonon coupling constant $\lambda(\omega)$ (dotted lines) for Lithium-doped
and Calcium-doped stanene.}
\label{LiCaEL}
\end{figure}

The Eliashberg spectral function $\alpha^2F(\omega)$  and the
electron-phonon coupling constant $\lambda(\omega)$ are plotted as
functions of the phonon energy $\omega$ in Fig. \ref{LiCaEL}.
Superconductivity in these doped stanene systems is clearly in the
intermediate coupling regime (i.e., $\lambda \sim 1$).
The contributions to the EPC $\lambda$  come
from three main ranges of phonon energies:
 the low energy phonons with energies from 0 to $\sim 60$~cm$^{-1}$, the
intermediate energy phonons ($\sim 80 - 150$~K) and the `high' energy
regime ($\gtrsim 150$~K). 
 For phonons with zero wavevector (${\bf
q}=0$), it is straightforward to determine which phonon modes correspond
to each of these regimes. However, for those with ${\bf q}\neq 0$, it is
difficult to unravel the phonon  
modes that contribute the most to a given phonon energy. Thus, it is
difficult to determine the modes that enhance $\lambda$ the most since
all the wavevectors in the first
Brillouin Zone contribute to the sum in Eq.~\ref{Eliashberg}. 
 The interaction between the electrons and the low
energy phonons contribute more than $70\%$ to the EPC. The remaining
$\sim 30\%$ comes mainly from the electronic coupling with the intermediate energy phonons. 
In general, we expect that out-of-plane
vibrational modes of stanene couples with the $\pi^*$ and interlayer
states to produce superconductivity in the doped system.

\subsection{Transition Temperature}
\begin{figure}[h!]
\begin{center}
\rotatebox{0}{\scalebox{0.5}{\includegraphics{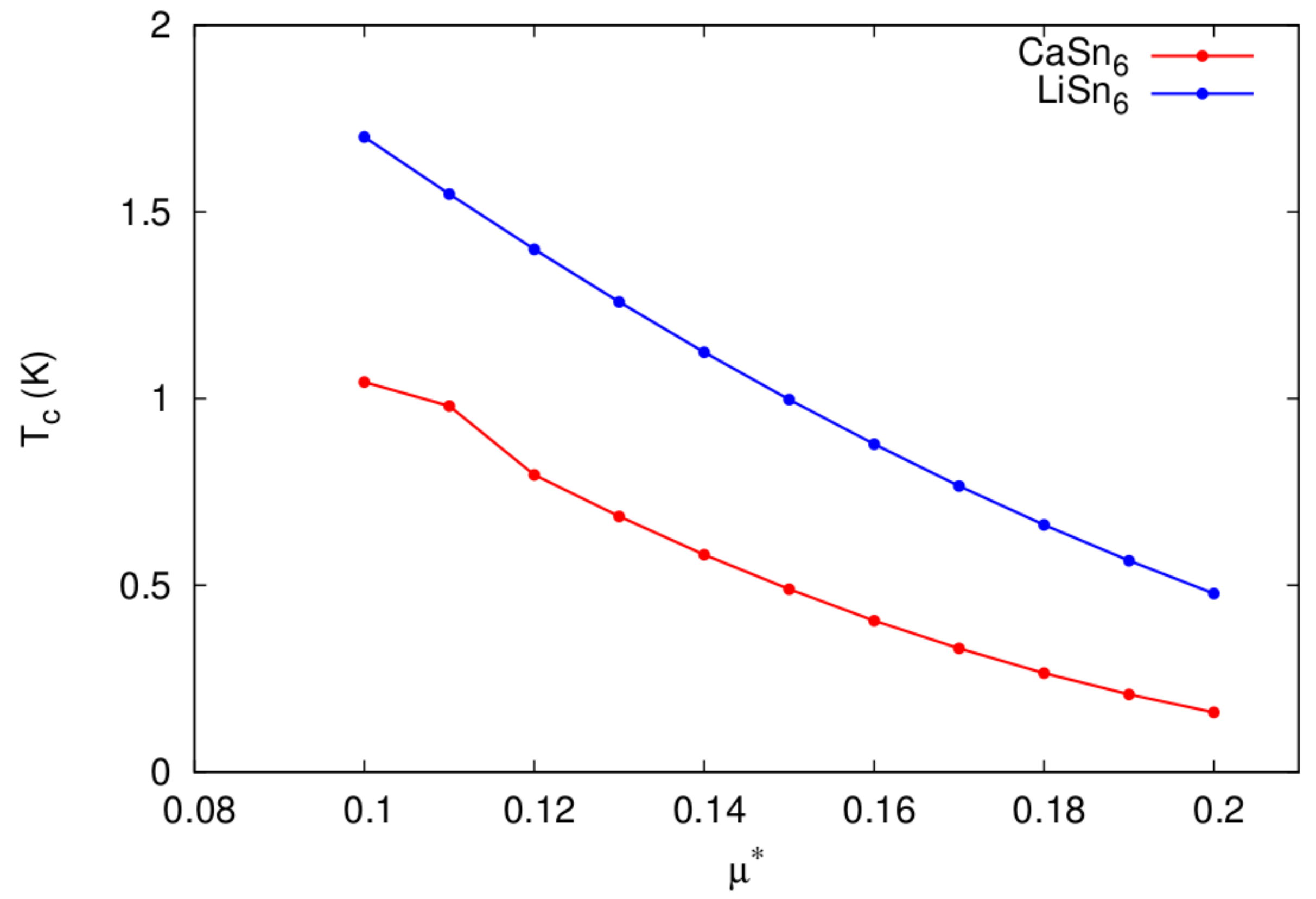}}}
\end{center}
\caption{Transition Temperature $T_c$ as a function of $\mu^*$} \label{TC}
\end{figure}

Table~\ref{RESULTS} shows the results obtained from the calculations. We
adopted a value of 
$\mu^*=0.13$ for the Coulombic repulsion parameter as this value reproduced, in a previous \ai\
calculation,~\cite{Sh14} the experimental $T_c$ of bulk white
Tin.
Our results follow the trend of transition temperatures $T_c$ observed by
Profeta \ea~\cite{PG12} with regard to the ad atoms on a graphene sheet.  
That is, the ad atom that gives the highest $T_c$ is the one whose
height above the plane is smallest. From Table~\ref{RESULTS},
we see that the height $\delta$ of the ad atom above the stanene plane
is smaller for Lithium doped stanene LiSn$_6$ than for its Calcium doped
counterpart CaSn$_6$ and, as expected, $T_c$ for the the former (1.3~K) is
higher than that of the latter (0.7~K). 
 One may argue that the value of the Coulomb repulsion parameter used in 
the bulk case might be different from that for the 2D system. In
Fig.~\ref{TC} we show the values of $T_c$ determined for different
values of $\mu^*$ in the range from 0.1 to 0.2 as suggested in
Ref.~\onlinecite{MC68}. The largest value of $T_c$ obtained was $1.7$~K
($1.0$~K) for LiSn$_6$ (CaSn$_6$). These values are significantly
smaller than that predicted for Lithium-doped graphene
($8.1$~K), but agree with the low $T_c$ value ($1.4$~K) for Calcium-doped graphene.~\cite{PG12}

In conclusion,  we  reported our results on phonon-mediated superconductivity in Lithium and Calcium doped stanene.
 We presented the electronic band structure, predicted the Eliashberg
spectral function $\alpha^2F(\omega)$, and showed the contribution of
the phonons to the total electron phonon coupling coefficient
$\lambda$. 
Our results reveal that superconductivity can be induced in stanene by doping it with alkaline and alkaline earth atoms. 
We found that $T_c$ for
Lithium doped stanene should lie in the range from $0.48$ to $1.70$~K
while that of the Calcium doped should lie somewhere from $0.15$ to
$1.04$~K. These values of $T_c$ are small, smaller than in the case for
bulk $\beta$-Tin and Lithium-doped graphene. Further work should seek
for methods to increase $T_c$ either by straining stanene or/and by
doping with other ad atoms. 

\section{Acknowledgments}
YS is grateful to the African Development Bank (AfDB) for a scholarship.
We thank the Abdus Salam International Center for Theoretical Physics
(ICTP), Trieste, Italy for computer resources.

\newpage
 \begin{center}
\begin{table}
 \caption{Optimized lattice parameter of stanene: $a$ is the lattice
constant, $d$ is the out-of-plane
buckled height, and $a_{sn-sn}$ is the $Sn-Sn$ bond length; all
distances are in Angstroms}\label{ST} 

\begin{tabular}{cccc} \hline \hline
              & $~~~~a~~~~$& $~~~~d~~~~$  &$~a_{sn-sn}~$\\ \hline
This work &4.64 & 0.92&2.84\\
Broek \ea$^{\cite{BB14}}$   & 4.62 & 0.92&2.82 \\
Shou-Cheng \ea$^{\cite{SZ13}}$&4.65&-&-\\ \hline  \hline
 \end{tabular}
\end{table}
 \end{center}

\newpage
\begin{center}
\begin{table}[h!]
 \caption{Summary of results: $\delta$ (in \AA) is the height of the ad
atom above the stanene plane, $\omega_{log}$ (in K) is the logarithmic
average of the
 phonon energy, $\lambda$ is the electron-phonon coupling coefficient, 
and $T_c$ (in K) is the transition temperature. }\label{RESULTS}
\begin{tabular}{ccccc} \hline \hline
    & $~~~~\delta~~~~$ &$~~~~\omega_{log}~~~~$& $~~~~\lambda~~~~$
&$~~~~T_c~~~~$\\ \hline 
 LiSn$_6$&1.75&60.97 & 0.65&1.3\\
 CaSn$_6$&2.24& 62.62& 0.54&0.7 \\
\hline  \hline
 \end{tabular}

\end{table}
 \end{center}

\newpage

\newpage

\newpage
\begin{figure}[h!]
\centering
\rotatebox{0}{\scalebox{0.5}{\includegraphics{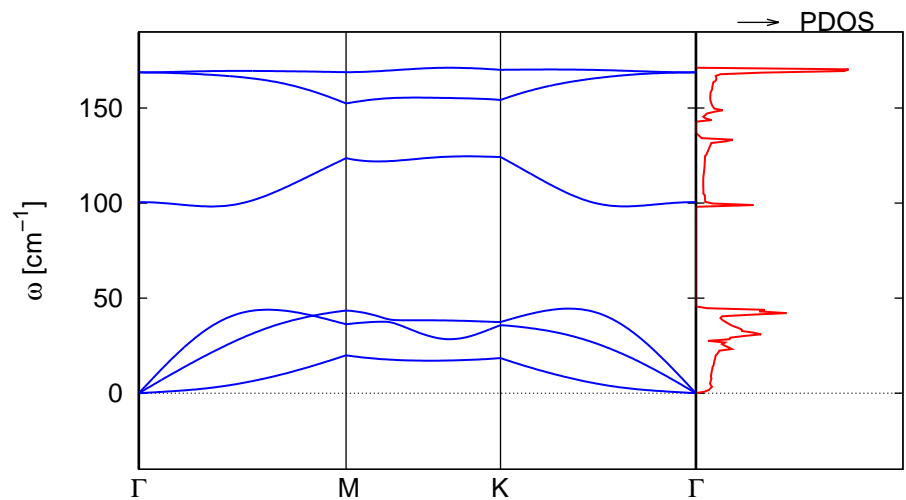}}}
\caption{Phonon dispersion (left) and the 
phonon density of states (right) for buckled pristine stanene. 
The phonon dispersion is plotted along $\Gamma,~M  ~K \text{and}~
\Gamma$ high symmetry k points of a 2D honeycomb crystal}\label{PS}
\end{figure}

\newpage

\newpage

\newpage
\newpage

\end{document}